\documentstyle[twocolumn,prl,epsfig,aps,floats]{revtex}
\tightenlines

\def\SrCu{Sr$_{14}$Cu$_{24}$O$_{41}$ }
\def\SrCax{Sr$_{14-x}$Ca$_x$Cu$_{24}$O$_{41}$ }
\def\SrCa12{Sr$_{2}$Ca$_{12}$Cu$_{24}$O$_{41}$ }
\def\Sr2.5{Sr$_{2.5}$Ca$_{11.5}$Cu$_{24}$O$_{41}$ }

\def\cm-1{cm$^{-1}$}

\begin{document}
\twocolumn[
\hsize\textwidth\columnwidth\hsize\csname@twocolumnfalse\endcsname
\draft
\title
{
Collective density wave excitations in two-leg
Sr$_{14-x}$Ca$_x$Cu$_{24}$O$_{41}$ ladders
}

\author{
A. Gozar$^{1,2}$, G.~Blumberg$^{1,\dag}$, P. B. Littlewood$^{3}$, B.S.
Dennis$^{1}$, N. Motoyama$^{4}$, H. Eisaki$^{5}$, and S. Uchida$^{4}$}

\address{
$^{1}$Bell Laboratories, Lucent Technologies, Murray Hill, NJ 07974 \\
$^{2}$University of Illinois at Urbana-Champaign, Urbana, IL
61801-3080 \\
$^{3}$University of Cambridge, Cavendish Laboratory, Cambridge, CB3
0HE UK\\
$^{4}$The University of Tokyo, Bunkyo-ku, Tokyo 113, Japan\\
$^{5}$Stanford University, Stanford, CA94305\\
}
\date{\today}
\maketitle

\begin{abstract}
Raman measurements in the $1.5 - 20$~\cm-1 energy range were performed
on single crystals of Sr$_{14-x}$Ca$_x$Cu$_{24}$O$_{41}$.
A quasielastic scattering peak (QEP) which softens with cooling is
observed in the polarization parallel to the ladder direction for samples
with $x = 0$, 8 and 12.
The QEP is a Raman fingerprint of pinned collective density
wave excitations screened by uncondensed carriers.
Our results suggest that transport in metallic
samples, which is similar to transport in
underdoped high-T$_{c}$ cuprates, is driven by a collective electronic
response.

\end{abstract}

]
\narrowtext

Competing ground states in low dimensional doped Mott-Hubbard systems
have been the subject of extensive research in recent years
\cite{Sachdev}.
Two-leg Cu-O based ladder materials like \SrCax
provide the opportunity to study not only magnetism in quasi
one-dimensional (1D) quantum systems but also charge carrier dynamics in 
an antiferromagnetic environment, with relevance to the phase diagram
of high-T$_{c}$ cuprates \cite{Dagotto}.
Magnetic correlations which give rise to a finite spin gap were
predicted to generate an attractive interaction between doped
carriers leading to superconductivity with a $d$-wave like order
parameter.
Due to the quasi-1D nature of these systems, ground states with
broken translational symmetry in which single holes or hole pairs can
order in a crystalline pattern are also possible.
The balance between superconducting and spin/charge density wave (DW)
ground states is ultimately determined by the microscopic parameters
of the theoretical models \cite{DagottoRice}.

The single crystals of \SrCax contain quasi-1D two-leg
Cu$_{2}$O$_{3}$ ladder planes which are stacked alternately with
planes of CuO$_{2}$ chains along the $b$ crystallographic axis
\cite{McCarronSiegrist}.
The ladder direction defines the $c$ axis and the lattice constants
of these two sub-systems satisfy
10$c_{chain}$~$\approx$~7$c_{ladder}$.
The nominal Cu valence in \SrCax is +2.25, independent of Ca
concentration.
In the insulating \SrCu crystals most of the carriers are believed to
be confined in the chains.
Transport and optical conductivity data
suggest that Ca substitution induces a transfer of holes from the
chains to the more conductive ladders \cite{Kato,Osafune97}.
The ladder carrier density was estimated from the optical spectral
weight to increase from 0.07 for $x = 0$ to about 0.2 for $x = 11$
\SrCax
\cite{Osafune97}.
A crossover to metallic conduction at high temperatures takes place
around $x = 11$ \cite{Osafune99} and for $x = 12$ the $c$-axis $dc$
resistivity has a minimum around T~=~70~K separating quasi-linear
metallic and insulating behavior similar to the case of high-T$_{c}$
cuprates in the underdoped regime~\cite{Vanacken,Motoyama,Ando}.
As opposed to \SrCax, the isostructural compound
La$_{6}$Ca$_8$Cu$_{24}$O$_{41}$ contains no holes per formula
unit.
At high Ca concentrations superconductivity under pressure has been
observed in \SrCax crystals with x~$\geq$~11.5 \cite{UeharaNagata}.

In the case of a DW instability, theory predicts the existence of
phase and amplitude collective modes of the DW order parameter
\cite{Lee}.
The amplitude excitation is Raman active and the phase mode should be
seen in optical absorption \cite{Lee,Gruner}.
In an ideal system the current carrying phase mode can slide
without friction \cite{Frohlich}, while impurities or lattice
commensurability destroy the infinite conductivity and shift this
mode to finite frequency as has been experimentally observed
\cite{Gruner,Degiorgi}.
In addition, many well established DW compounds display a loss peak
that has strongly temperature dependent energy and damping relating to
the $dc$ conductivity of the material \cite{Relaxation}.
This screened longitudinal excitation has been observed in the
transverse
response by measurements of the complex finite frequency dielectric
constant $\epsilon(\omega)$.
Electronic Raman scattering can probe the longitudinal electronic
channel, essentially the response of the charge density because the
Raman response function $\chi''(\omega) \propto
Im(1/\epsilon(\omega))$~\cite{Miles}.
The existence of collective DW excitations is clearly established for
\SrCu \cite{GirshScience} by measurements of non-linear
conduction and the relaxational dielectric response in the
10~-~10$^{6}$ Hz~region displaying a scattering rate that
scales with the $dc$ conductivity. 

One important question is what are the Raman signatures of these
collective modes and whether DW correlations still persist at higher
carrier dopings in ladder systems.
Here we present low frequency Raman scattering results that reveal
longitudinal (screened) collective charge density oscillations between
250 and 650~K in \SrCax crystals within a wide concentration range,
$0 < x < 12$.
The characteristic quasi-elastic scattering peak (QEP) we observed in
the
1.5~-~8~\cm-1 range above 300~K softens with cooling and is present
only for  polarization parallel the ladder direction.
A hydrodynamic model \cite{Littlewood} quantitatively accounts for
the collective excitations seen in the Raman response for \SrCu
compound.
The presence of the QEP in \SrCa12 demonstrates that density wave
correlations are present, at least for temperatures above 250~K, even
in superconducting (under pressure) crystals.

We measured Raman scattering from freshly cleaved $ac$ surfaces of
\SrCax and La$_{6}$Ca$_8$Cu$_{24}$O$_{41}$ single crystals grown as
described in \cite{Osafune97,Motoyama}.
Excitation energies of 1.55 and 1.65~eV from a Kr$^{+}$ laser were
used.
The spectra were taken using a custom triple grating
spectrometer and corrected for the spectral response of the
spectrometer and detector.
For measurements below T~=~300~K the samples were mounted in a
continuous flow optical cryostat and for above room
temperature in a TS1500 Linkam heat stage.
Stokes and anti-Stokes spectra were taken
for all data above 300~K to determine the temperature in the laser
spot.

Fig.~1 shows temperature dependent Raman spectra for \SrCu and \SrCa12 
in $cc$ polarization.
The low frequency Raman spectra at high temperatures in both crystals
look qualitatively similar.
They are dominated by the presence of strong quasi-elastic scattering
rising from the lowest measured energy of about 1.5~\cm-1 and peaked
about 7~\cm-1 for temperatures around 620~K.
With cooling the QEP shifts to lower frequencies and it gains
spectral weight.
Below $T \approx 450$~K the peak position moves below the instrumental 
cut-off energy and only the high frequency tail of the peak is observed. 
\begin{figure}[t]
\centerline{
\epsfig{figure=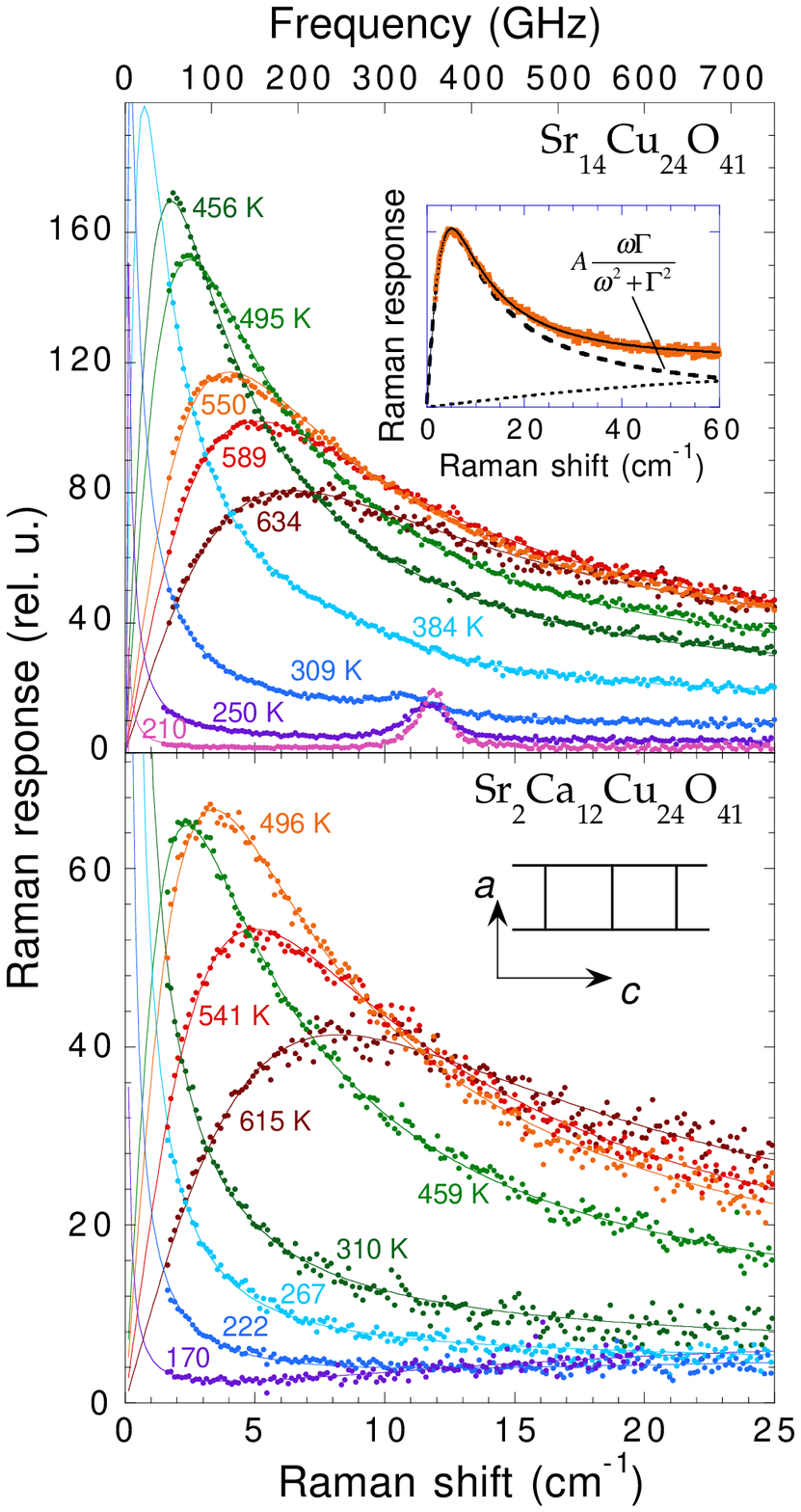,width=70mm}
}
\caption{Temperature dependent Raman response for $cc$ polarization in 
\SrCu and \SrCa12 taken with the 1.55~eV excitation energy.
Upper inset: Typical fit of the Raman data with a
relaxational form, Eq.~(\ref{one}), and a small contribution from an
underlying background.
For temperatures below 310~K the fits for \SrCu included also
the phonon around 12~\cm-1.
Lower inset: Two-leg ladder structure and axes
notation.}
\label{Fig.1}
\end{figure}

The polarization dependence of the QEP are summarized in Fig.~2.
For \SrCu (Fig.~2a) the QEP is present in $cc$ and absent in $aa$
polarization.
We do not observe the QEP in $cc$ polarization for
La$_{6}$Ca$_8$Cu$_{24}$O$_{41}$ (Fig.~2b) which contains no holes
per formula unit.
The presence of quasi-elastic scattering for $x > 0$  \SrCax
exhibiting the
same polarization selection rules as
shown in panels~$c$ and $d$ of Fig.~2 proves that this feature is a
characteristic of these compounds at all Ca substitution levels.
Applied magnetic fields up to 8~T influenced neither the energy of
the QEP nor the modes seen in Fig.~2a and b at 12~\cm-1 in \SrCu and
about 15~\cm-1 in La$_{6}$Ca$_{8}$Cu$_{24}$O$_{41}$.
We can conclude that the latter features are phonons.
The unusually low energy of these modes which points
towards a very high effective mass oscillator is interesting.
These  'folded' phonons appear as a result of chain-ladder
incommensurability \cite{McCarronSiegrist} that gives rise to a big
unit cell.

\begin{figure}[t]
\centerline{
\epsfig{figure=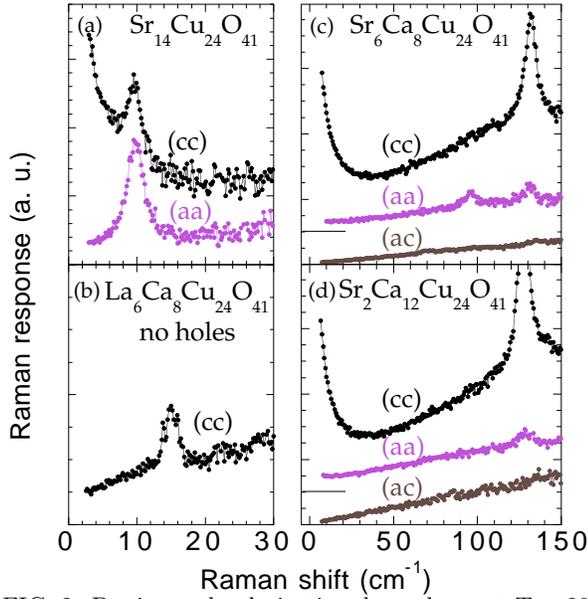,width=80mm}
}
\caption{Doping and polarization dependence at T~=~295~K of the 
quasielastic 
scattering peak (QEP) in spectra taken with 1.65~eV excitation.
(a) For \SrCu the QEP is present in $cc$ and absent in $aa$
polarizations.
(b) For La$_{6}$Ca$_8$Cu$_{24}$O$_{41}$ the QEP is not present.
In panels (c) and (d) we observe the QEP for $x = 8$
and 12 Sr$_{14-x}$Ca$_x$Cu$_{24}$O$_{41}$ only in
$cc$ polarization. The $aa$ data is offset.}
\label{Fig.2}
\end{figure}
The inset in Fig.~1 shows a typical deconvolution of the Raman data
by a fit to a relaxational form:
\begin{equation}
\chi'' (\omega) = A(T) \frac{\omega \Gamma}{\omega^2 + \Gamma^2}.
\label{one}
\end{equation}
A second phenomenological term accounting for a small underlying
electronic background was also used.
The temperature dependent fitting parameters are shown in the Fig.~3.
An Arrhenius plot for the damping parameter $\Gamma$ reveals an
activated behavior of the form $\Gamma (T) \propto$~ exp~$( - \Delta / 
T)$ with $\Delta \approx$~2100~K and 2180~K for \SrCu and \SrCa12 
respectively. 
These energies are very close to the activation energy displayed by
the $c$~axis $dc$ conductivity for \SrCu crystal
\cite{Motoyama,McElfresh} shown in the inset of Fig.~3.
The Raman QEP traces to high temperatures
the relaxational peak seen in transport \cite{GirshScience},
suggesting their common origin.
Also, the frequency of the QEP, which is much lower than the thermal
energy, the magnetic or $dc$ activation gaps, points
toward a collective rather than a single particle excitation.
For \SrCa12 the QEP is also present (Fig.~1b).
The relaxation parameter $\Gamma$ is slightly higher than for
\SrCu sample while the spectral weight is lower. 
However, for \SrCa12 the c-axis $dc$ conductivity above
T~$\approx$~70~K shows a metallic behavior.
\begin{figure}[t]
\centerline{
\epsfig{figure=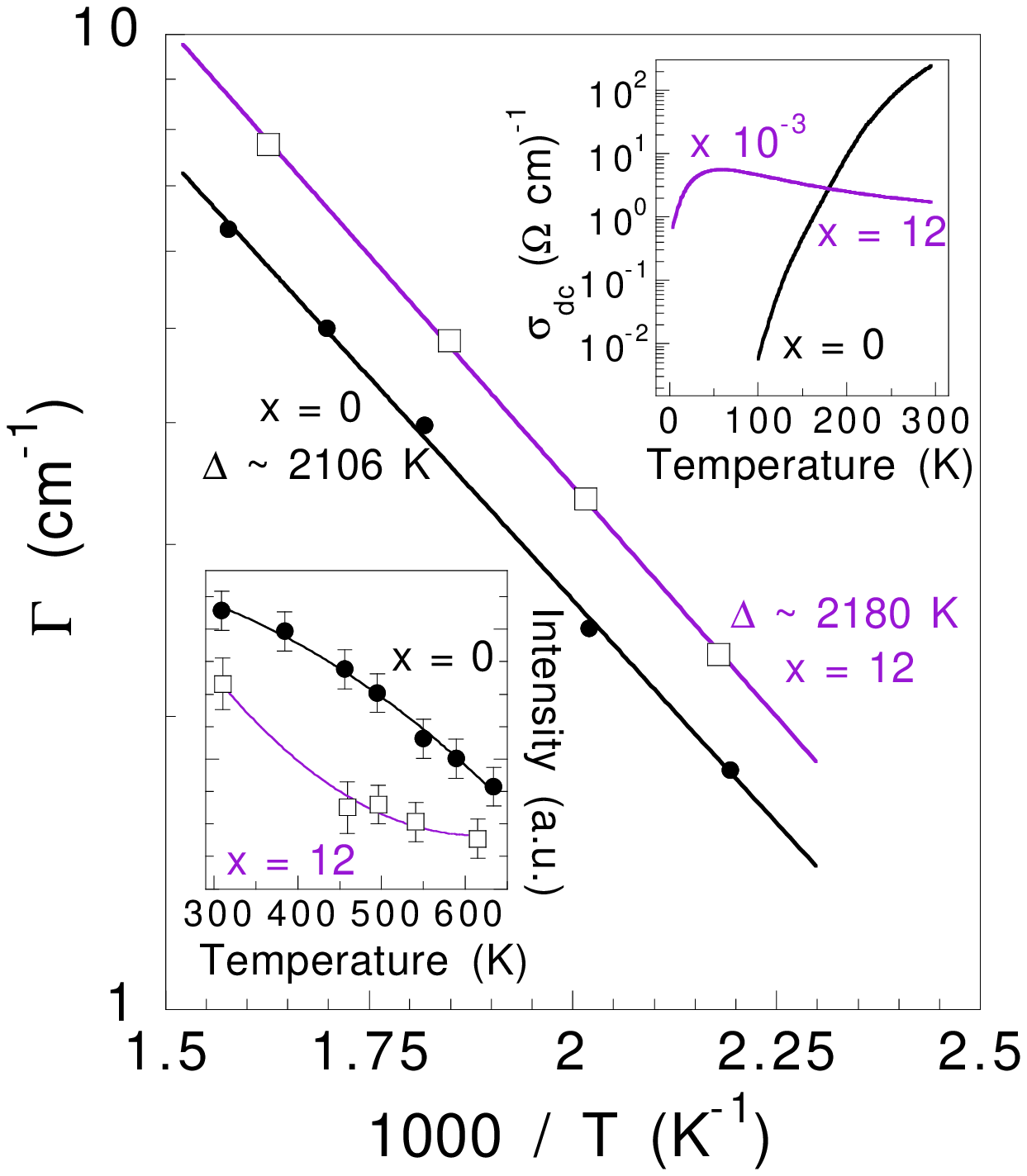,width=64mm}
}
\caption{Temperature dependence of the relaxational rate $\Gamma$ for 
\SrCu (filled circles) and \SrCa12 (empty squares).
An activated behavior $\Gamma \propto exp(-\Delta/T)$ is observed
for both compounds.
Upper inset: Temperature dependence of the $dc$ conductivity as a
function of temperature.
Note that the data for \SrCa12 was divided by a factor of 10$^{3}$.
Lower inset: Variation of the quasi-elastic peak weight, A(T),
with temperature (solid lines are guides for the eye).}
\label{Fig.3}
\end{figure}

In the analysis of the Raman spectra shown in Fig.~1 we interpret the 
low frequency overdamped excitation as a
DW relaxational mode in the longitudinal channel.
The interaction of longitudinal DW modes with normal (uncondensed)
carriers resembles to some degree the problem of coupling of
plasma oscillations to the longitudinal optical (LO) vibrations in
doped semiconductors~\cite{Yokota}.
The longitudinal modes of one excitation interact with the
electrostatic field
produced by the other and as result their bare energy gets
renormalized due to screening effects.
Essentially a feature which should be observed only in the
longitudinal channel, the DW mode leaks into the transverse response
due to the non-uniform pinning which introduces disorder mixing the
pure transverse
and longitudinal character of the excitations \cite{Littlewood}.
This was seen for \SrCu \cite{GirshScience} as well as for
other DW compounds \cite{Relaxation}.
Modelling the DW contribution to the dielectric function by an
oscillator, we have:
\begin{equation}
\epsilon_{DW} (\omega) = \frac{\Omega_{p}^{2}}{\omega^2 - i
\gamma_{0} \omega - \Omega_{0}^{2}} .
\label{two}
\end{equation}
where $\Omega_{0}$ is the characteristic pinning frequency,
$\gamma_{0}$ an
intrinsic damping coefficient and $\Omega_{p}^{2} = 4 \pi
\rho_{c}^{2} / m^{*}$ is the spectral weight of the DW given in terms
of $\rho_{c}$ and $m^{*}$ which are the associated charge and
effective mass densities.
A more realistic model would allow for a broad distribution of
pinning frequencies, extending over several decades, as inferred from
the low frequency response\cite{Relaxation,GirshScience}.
Within a two-fluid model, assuming that the only coupling between
condensed and normal electrons is by an electric field due to density
fluctuations, the longitudinal response which accounts for the
screening effects and is relevant for Raman scattering can be written
as~\cite{Littlewood}:
\begin{equation}
\epsilon_{L} = \frac{\Omega_{p}^{2}}{\Omega_{0}^{2} - \omega^2 - i
\gamma_{0} \omega - \frac{i \omega \Omega_{p}^{2}}{4 \pi \sigma_{dc}
- i \omega \epsilon_{\infty}}}
\label{three}
\end{equation}
Here $\sigma_{dc}$ is the 'quasiparticle' conductivity and
$\epsilon_{\infty}$ takes into account the contribution of high
frequency interband transitions to the polarization.
Note that in the low frequency limit, Eq.~(\ref{three}) reduces to
the relaxational form of Eq.~(\ref{one}), with a weight A(T) =
$\Omega_{p}^{2} / \Omega_{0}^{2}$ and a damping
\begin{equation}
\Gamma$~(T) = 4$\pi \sigma_{dc} \Omega_{0}^{2} / \Omega_{p}^{2}.
\label{four}
\end{equation}
The proportionality between $\Gamma$ and the the $dc$ conductivity,
which is clearly seen in Fig.~3 for \SrCu, is the
result of normal carrier backflow which screens the collective
polarization and dissipates energy, suffering lattice momentum
relaxation \cite{Littlewood}.

Extrapolating the activated fit for the conductivity in between 150
and 300~K for \SrCu to temperatures higher than 450~K, where the
QEP was still seen in Raman spectra, we calculated the values for
$\Gamma(T)$ from Eq.~(\ref{four}) using $\Omega_{0} \approx$
1-4~\cm-1 and $\Omega_{p} \approx$ 3300~\cm-1 as determined from low
temperature microwave data \cite{Kitano} and the $c$-axis loss
function \cite{Eisaki}.
This independent calculation yields values which are lower by a factor
of 25 than the Raman data suggests.
There are two reasons which can explain this discrepancy.
One is related to the reduction in the density of condensed carriers
in close proximity to the DW transition.
The second is that the distribution of pinning frequencies is in the
range of measured Raman shifts, so that we obtain a contribution to a
Raman response of the form of Eq.~(\ref{one}) only from the higher
part of the distribution.
Furthermore, above T~$\approx$~150~K the width of the pinning
frequencies distribution increases.
Although $\Omega_{0}$ might still be centered around 1-4~\cm-1
\cite{Kitano},
the mode becomes strongly broadened at high temperatures.
Eq.~(\ref{three}) also predicts a charge DW plasmon at the frequency
$(\Omega_{0}^{2} + \Omega_{p}^{2} / \epsilon_{\infty})^{1/2}\approx$
1500~\cm-1.
We do not observe this feature in our spectra due to the strong
damping and the fact that it lies in the midst of a strong
multi-phonon scattering region.

For \SrCa12 above 70~K the $dc$ conductivity is metallic
\cite{Motoyama}.
However, the Raman response in the 2~-~8~\cm-1 region
(Fig.~1) is qualitatively similar to the \SrCu crystal.
The similarity of the results allows us to conclude that DW
correlations are also present at high Ca substitution levels.
We propose that the metallic behavior for \SrCa12 is due to a
partially
gapped Fermi surface.
Support for this conjecture comes from an angle resolved
photoemission study \cite{Takahashi} which shows that while for \SrCu
the gap is finite, for Sr$_{5}$Ca$_9$Cu$_{24}$O$_{41}$  the density of
states rises almost to the chemical potential.
This spectral weight transfer is enhanced with further increase in the 
ladder hole concentration \cite{Osafune97}.
The observation of an activated relaxation rate in the metallic
x~=~12 compound is not consistent with Eq.~(3), based on a simple
two-fluid assumption.
Note however that by room temperature the x~=~0 and x~=~12 compounds
have comparable conductivities, suggesting that the
contribution of the remnant Fermi surface to the overall carrier
density is quite small.
Similar relaxation rates for \SrCu and \SrCa12 might be
reconciled with different transport properties assuming a strongly
momentum dependent scattering rate.
Carrier condensation in the DW state leads to a completely
gapped Fermi surface resulting in an insulating behavior below
T~=~70~K.
In addition, the DW dynamics in \SrCa12 is influenced by the
random potential introduced in the system because of cation
substitution which affects the pinning mechanism \cite{Schneemeyer}.
Another more speculative explanation for the metallic like
conductivity in  \SrCa12 could be that in the presence of a very
broad distribution of pinning frequencies $\Omega_{0}$ as a result of
Ca substitution, not all the collective contribution gets pinned, and
we still have a Fr\"{o}hlich type component \cite{Frohlich}
contributing to the $dc$ conductivity.
Irrespective of the exact microscopic model, strong similarities
between local structural units and transport
properties in Cu-O based ladders and underdoped high T$_{c}$
materials suggest that carrier dynamics in 2D Cu-O
sheets at low hole concentration could be also governed by a collective
DW response.

In conclusion we demonstrated the existence of DW correlations in
doped two-leg \SrCax ladders.
We found Raman fingerprints of screened longitudinal collective modes
in crystals with Ca concentrations from $x = 0$ to 12.
A hydrodynamic model was used to quantitatively account for the
existence of the charge collective mode in the insulating
Sr$_{14}$Cu$_{24}$O$_{41}$ compound whose damping scales with the 
activated conductivity. 
This mode is also present in the superconducting (under
pressure) Sr$_{2}$Ca$_{12}$Cu$_{24}$O$_{41}$ ladder.
Our results demonstrate that the paired superconducting state competes 
in these materials with a crystalline charge ordered ground
state.


\begin{references}

\vspace{-15mm}

\bibitem[\dag]{byline}  To whom correspondence should be addressed.
E-mail: girsh@bell-labs.com

\bibitem{Sachdev}
S. Sachdev, Science {\bf 288}, 475 (2000).

\bibitem{Dagotto}
E. Dagotto and T. M. Rice, Science {\bf 271}, 618 (1996); E. Dagotto,
Rep. Prog. Phys. {\bf 62}, 1525 (1999) and references
therein.

\bibitem{DagottoRice}
E. Dagotto , J. Riera, and D. Scalapino, Phys. Rev. B {\bf 45}, 5744
(1992); T. M. Rice, S. Gopalan, and M. Sigrist, Europhys. Lett. {\bf
23}, 445 (1993).

\bibitem{McCarronSiegrist}
E. M. McCarron III {\em et al.}, Mater. Res. Bull. {\bf 23}, 1355
(1988); T. Siegrist {\em et al.}, Mater. Res. Bull. {\bf 23}, 1429
(1988).

\bibitem{Kato}
M. Kato {\em et al.},  Physica C {\bf 258}, 284 (1996).

\bibitem{Osafune97}
T. Osafune {\em et al.}, Phys. Rev. Lett. {\bf 78}, 1980 (1997).

\bibitem{Osafune99}
T. Osafune {\em et al.}, Phys. Rev. Lett. {\bf 82}, 1313 (1999).

\bibitem{Vanacken}
J. Vanacken {\em et al.},  Physica C {\bf 337}, 260 (2000).

\bibitem{Motoyama}
N. Motoyama {\em et al.}, Phys. Rev. B {\bf 55}, R3386 (1997).

\bibitem{Ando}
Y. Ando {\em et al.}, Phys. Rev. Lett. {\bf 87}, 017001 (2001).

\bibitem{UeharaNagata}
M. Uehara {\em et al.}, J. Phys. Soc. Jpn. {\bf 65}, 2764 (1996); T.
Nagata {\em et al.}, Phys. Rev. Lett. {\bf 81}, 1090 (1998).

\bibitem{Lee}
P. A. Lee, T. M. Rice, and P. W. Anderson, Solid State Comm. {\bf
14}, 703 (1974).

\bibitem{Gruner}
G. Gr\"{u}ner, Density waves in solids (Perseus publishing,
Cambridge, Massachusetts, 1994).

\bibitem{Frohlich}
F. R. S. Fr\"{o}hlich, Proc. R. Soc., Lond. {\bf A223}, 296 (1954).

\bibitem{Degiorgi}
L. Degiorgi {\em et al.}, Phys. Rev. B {\bf 44}, 7808 (1991) and
references therein.

\bibitem{Relaxation}
W. Wu {\em et al.}, Phys. Rev. Lett. {\bf 52}, 2382 (1984); R. J.
Cava {\em et al.}, Phys. Rev. B {\bf 33}, 2439 (1986); {\bf 30}, 3228
(1984).

\bibitem{Miles}
M. V. Klein, Chap. 4 in Light Scattering in Solids I, (Ed. M.
Cardona, Springer-Verlag, 1983)

\bibitem{GirshScience}
G. Blumberg {\em et al.}, Science {\bf 297}, 26 July (2002).

\bibitem{Littlewood}
P. B. Littlewood, Phys. Rev. B {\bf 36}, 3108 (1987).

\bibitem{McElfresh}
M. W. McElfresh {\em et al.}, Phys. Rev. B {\bf 40}, 825 (1989).

\bibitem{Yokota}
I. Yokota, J. Phys. Soc. Jpn. {\bf 16}, 2075 (1961).

\bibitem{Kitano}
H. Kitano {\em et al.}, Europhys. Lett. {\bf 56}, 434 (2001).

\bibitem{Eisaki}
H. Eisaki {\em et al.},  Physica C {\bf 341-348}, 363 (2000).

\bibitem{Schneemeyer}
L. F. Schneemeyer {\em et al.}, Phys. Rev. B {\bf 30}, 4297 (1984).

\bibitem{Takahashi}
T. Takahashi {\em et al.}, Phys. Rev. B {\bf 56}, 7870 (1997).

\end{references}
\end{document}